\RequirePackage{lineno}
\documentclass[prl,twocolumn,superscriptaddress]{revtex4}
\usepackage{color}

\usepackage{amsmath}
\usepackage{graphicx}
\usepackage{url}
\usepackage{color}
\begin{document}
\setpagewiselinenumbers
%

\title{ Measurement of Exclusive $\pi^0$ Electroproduction Structure Functions and their Relationship to   Transversity GPDs}

\newcommand*{\ANL}{Argonne National Laboratory, Argonne, Illinois 60439}
\newcommand*{\ANLindex}{1}
\affiliation{\ANL}
\newcommand*{\ASU}{Arizona State University, Tempe, Arizona 85287-1504}
\newcommand*{\ASUindex}{2}
\affiliation{\ASU}
\newcommand*{\CSUDH}{California State University, Dominguez Hills, Carson, CA 90747}
\newcommand*{\CSUDHindex}{3}
\affiliation{\CSUDH}
\newcommand*{\CANISIUS}{Canisius College, Buffalo, NY}
\newcommand*{\CANISIUSindex}{4}
\affiliation{\CANISIUS}
\newcommand*{\CMU}{Carnegie Mellon University, Pittsburgh, Pennsylvania 15213}
\newcommand*{\CMUindex}{5}
\affiliation{\CMU}
\newcommand*{\CUA}{Catholic University of America, Washington, D.C. 20064}
\newcommand*{\CUAindex}{6}
\affiliation{\CUA}
\newcommand*{\SACLAY}{CEA, Centre de Saclay, Irfu/Service de Physique Nucl\'eaire, 91191 Gif-sur-Yvette, France}
\newcommand*{\SACLAYindex}{7}
\affiliation{\SACLAY}
\newcommand*{\CNU}{Christopher Newport University, Newport News, Virginia 23606}
\newcommand*{\CNUindex}{8}
\affiliation{\CNU}
\newcommand*{\UCONN}{University of Connecticut, Storrs, Connecticut 06269}
\newcommand*{\UCONNindex}{9}
\affiliation{\UCONN}
\newcommand*{\EDINBURGH}{Edinburgh University, Edinburgh EH9 3JZ, United Kingdom}
\newcommand*{\EDINBURGHindex}{10}
\affiliation{\EDINBURGH}
\newcommand*{\FU}{Fairfield University, Fairfield CT 06824}
\newcommand*{\FUindex}{11}
\affiliation{\FU}
\newcommand*{\FIU}{Florida International University, Miami, Florida 33199}
\newcommand*{\FIUindex}{12}
\affiliation{\FIU}
\newcommand*{\FSU}{Florida State University, Tallahassee, Florida 32306}
\newcommand*{\FSUindex}{13}
\affiliation{\FSU}
\newcommand*{\Genova}{Universit$\grave{a}$ di Genova, 16146 Genova, Italy}
\newcommand*{\Genovaindex}{14}
\affiliation{\Genova}
\newcommand*{\GWUI}{The George Washington University, Washington, DC 20052}
\newcommand*{\GWUIindex}{15}
\affiliation{\GWUI}
\newcommand*{\ISU}{Idaho State University, Pocatello, Idaho 83209}
\newcommand*{\ISUindex}{16}
\affiliation{\ISU}
\newcommand*{\INFNFE}{INFN, Sezione di Ferrara, 44100 Ferrara, Italy}
\newcommand*{\INFNFEindex}{17}
\affiliation{\INFNFE}
\newcommand*{\INFNFR}{INFN, Laboratori Nazionali di Frascati, 00044 Frascati, Italy}
\newcommand*{\INFNFRindex}{18}
\affiliation{\INFNFR}
\newcommand*{\INFNGE}{INFN, Sezione di Genova, 16146 Genova, Italy}
\newcommand*{\INFNGEindex}{19}
\affiliation{\INFNGE}
\newcommand*{\INFNRO}{INFN, Sezione di Roma Tor Vergata, 00133 Rome, Italy}
\newcommand*{\INFNROindex}{20}
\affiliation{\INFNRO}
\newcommand*{\ORSAY}{Institut de Physique Nucl\'eaire ORSAY, Orsay, France}
\newcommand*{\ORSAYindex}{21}
\affiliation{\ORSAY}
\newcommand*{\ITEP}{Institute of Theoretical and Experimental Physics, Moscow, 117259, Russia}
\newcommand*{\ITEPindex}{22}
\affiliation{\ITEP}
\newcommand*{\JMU}{James Madison University, Harrisonburg, Virginia 22807}
\newcommand*{\JMUindex}{23}
\affiliation{\JMU}
\newcommand*{\KNU}{Kyungpook National University, Daegu 702-701, Republic of Korea}
\newcommand*{\KNUindex}{24}
\affiliation{\KNU}
\newcommand*{\LPSC}{LPSC, Universite Joseph Fourier, CNRS/IN2P3, INPG, Grenoble, France
}
\newcommand*{\LPSCindex}{25}
\affiliation{\LPSC}
\newcommand*{\UNH}{University of New Hampshire, Durham, New Hampshire 03824-3568}
\newcommand*{\UNHindex}{26}
\affiliation{\UNH}
\newcommand*{\NSU}{Norfolk State University, Norfolk, Virginia 23504}
\newcommand*{\NSUindex}{27}
\affiliation{\NSU}
\newcommand*{\OHIOU}{Ohio University, Athens, Ohio  45701}
\newcommand*{\OHIOUindex}{28}
\affiliation{\OHIOU}
\newcommand*{\ODU}{Old Dominion University, Norfolk, Virginia 23529}
\newcommand*{\ODUindex}{29}
\affiliation{\ODU}
\newcommand*{\RPI}{Rensselaer Polytechnic Institute, Troy, New York 12180-3590}
\newcommand*{\RPIindex}{30}
\affiliation{\RPI}
\newcommand*{\URICH}{University of Richmond, Richmond, Virginia 23173}
\newcommand*{\URICHindex}{31}
\affiliation{\URICH}
\newcommand*{\ROMAII}{Universita' di Roma Tor Vergata, 00133 Rome Italy}
\newcommand*{\ROMAIIindex}{32}
\affiliation{\ROMAII}
\newcommand*{\MSU}{Skobeltsyn Nuclear Physics Institute, 119899 Moscow, Russia}
\newcommand*{\MSUindex}{33}
\affiliation{\MSU}
\newcommand*{\SCAROLINA}{University of South Carolina, Columbia, South Carolina 29208}
\newcommand*{\SCAROLINAindex}{34}
\affiliation{\SCAROLINA}
\newcommand*{\JLAB}{Thomas Jefferson National Accelerator Facility, Newport News, Virginia 23606}
\newcommand*{\JLABindex}{35}
\affiliation{\JLAB}
\newcommand*{\UNIONC}{Union College, Schenectady, NY 12308}
\newcommand*{\UNIONCindex}{36}
\affiliation{\UNIONC}
\newcommand*{\UTFSM}{Universidad T\'{e}cnica Federico Santa Mar\'{i}a, Casilla 110-V Valpara\'{i}so, Chile}
\newcommand*{\UTFSMindex}{37}
\affiliation{\UTFSM}
\newcommand*{\GLASGOW}{University of Glasgow, Glasgow G12 8QQ, United Kingdom}
\newcommand*{\GLASGOWindex}{38}
\affiliation{\GLASGOW}
\newcommand*{\VIRGINIA}{University of Virginia, Charlottesville, Virginia 22901}
\newcommand*{\VIRGINIAindex}{39}
\affiliation{\VIRGINIA}
\newcommand*{\WM}{College of William and Mary, Williamsburg, Virginia 23187-8795}
\newcommand*{\WMindex}{40}
\affiliation{\WM}
\newcommand*{\YEREVAN}{Yerevan Physics Institute, 375036 Yerevan, Armenia}
\newcommand*{\YEREVANindex}{41}
\affiliation{\YEREVAN}

\newcommand*{\NOWMSU}{Skobeltsyn Nuclear Physics Institute, 119899 Moscow, Russia}
\newcommand*{\NOWORSAY}{Institut de Physique Nucl\'eaire ORSAY, Orsay, France}
\newcommand*{\NOWINFNGE}{INFN, Sezione di Genova, 16146 Genova, Italy}
\newcommand*{\NOWROMAII}{Universita' di Roma Tor Vergata, 00133 Rome Italy}

\author {I.~Bedlinskiy}
\affiliation{\ITEP}
\author {V.~Kubarovsky} 
\affiliation{\JLAB}
\affiliation{\RPI}
\author {S.~Niccolai}
\affiliation{\ORSAY}
\author {P.~Stoler} 
\affiliation{\RPI}

\author {K.P. ~Adhikari} 
\affiliation{\ODU}
\author {M.~Aghasyan}
\affiliation{\INFNFR}
\author {M.J.~Amaryan} 
\affiliation{\ODU}
\author {M.~Anghinolfi} 
\affiliation{\INFNGE}
\author {H.~Avakian}
\affiliation{\JLAB}
\author {H.~Baghdasaryan} 
\affiliation{\VIRGINIA}
\affiliation{\YEREVAN}
% ...

\collaboration{The CLAS Collaboration}
\noaffiliation

%
%
%
%
%
%
%

\date{\today}

\begin{abstract}
Exclusive $\pi^0$ electroproduction at a  beam energy of 5.75 GeV has been measured  with 
the Jefferson Lab CLAS spectrometer. Differential cross sections were measured at more than 1800 kinematic values in $Q^2$, $x_B$, $t$, and $\phi_\pi$,
in the  $Q^2$  range from 1.0 to 4.6 GeV$^2$,\  $-t$ up to  2 GeV$^2$,  and $x_B$ from 0.1 to 0.58.  Structure functions $\sigma_T +\epsilon \sigma_L,  \sigma_{TT}$ and $\sigma_{LT}$ were extracted as functions of $t$ for each of  17 combinations of  $Q^2$ and $x_B$.  The data were compared directly with two handbag-based  calculations including both  longitudinal and transversity GPDs. Inclusion of only longitudinal GPDs very strongly underestimates $\sigma_T +\epsilon \sigma_L$ and fails to account for $\sigma_{TT}$ and  $\sigma_{LT}$, while inclusion of transversity GPDs brings the calculations into substantially better agreement with the data. There is very strong sensitivity to the relative contributions of nucleon helicity flip and helicity non-flip processes.
The results  confirm  that exclusive $\pi^0$ electroproduction  offers direct experimental access to the  transversity GPDs. 
\\
\\*DOI: 11.1103/PhysRevLett.109.112001 \hfill PACS numbers: 13.60Le, 14.20.Dh, 14.40.Be, 24.85.+p

\end{abstract}

\maketitle

A  major goal of hadronic physics is to describe the three dimensional structure of the nucleon in terms of its quark and gluon fields. Deep inelastic scattering experiments  have provided a large body of information about quark longitudinal momentum distributions.  
Exclusive electron scattering experiments, in which all final state particles are measured, have been rather successfully analyzed and interpreted by Regge models which are based on hadronic degrees of freedom (see, for example, Refs.~\cite{laget,kaskulov}).
 
\begin{figure}[h]
\includegraphics[width=3.0in]{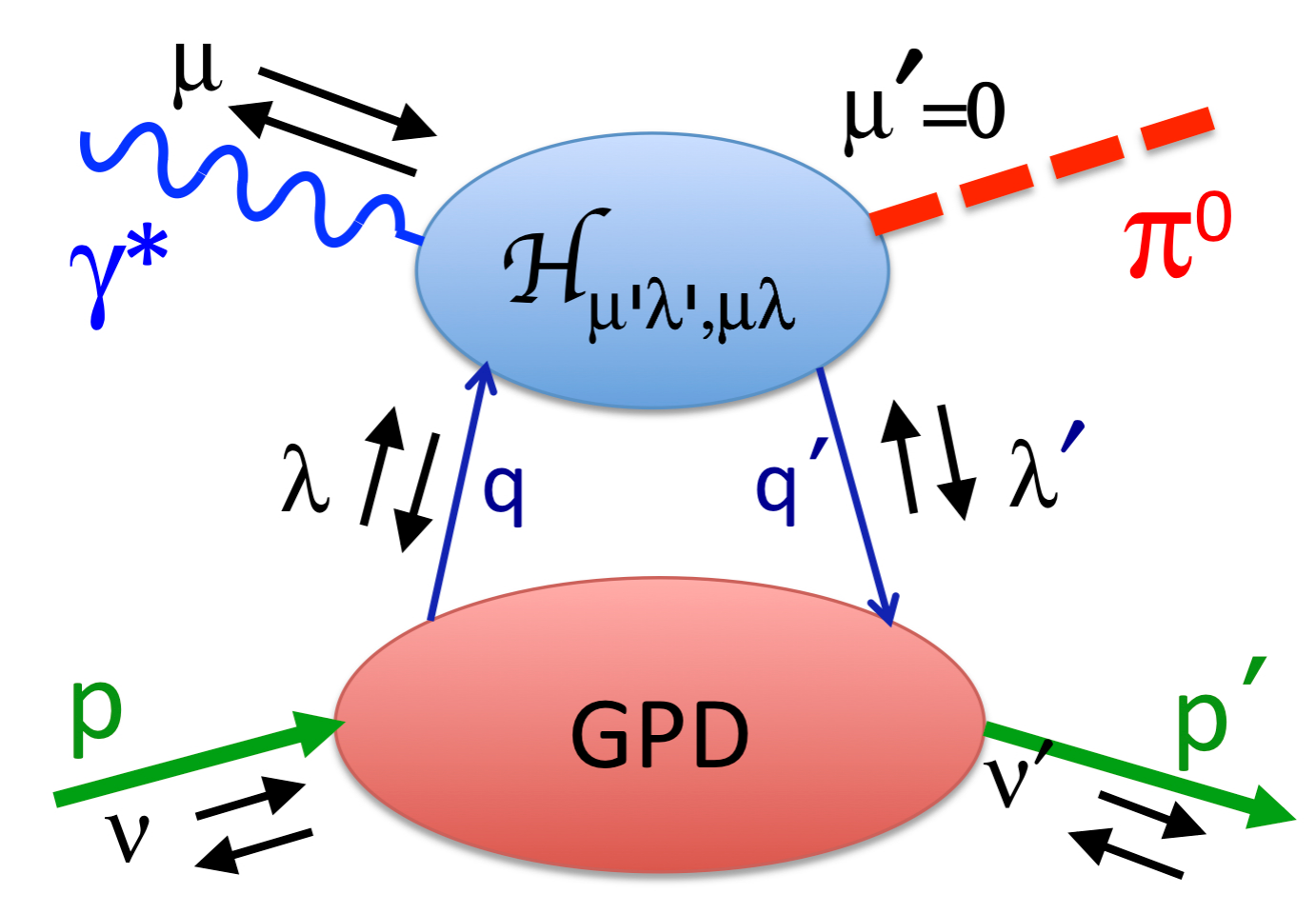}
\caption{\label{fig:handbag} Schematic diagram of the $\pi^0$ electroproduction amplitude in the framework of  the handbag mechanism.  The helicities of the initial and final nucleons are denoted by $\nu$ and $\nu^\prime$, the incident photon and produced meson by $\mu$ and $\mu^\prime$ and the active initial  and final quark by $\lambda$ and $\lambda^\prime$. The arrows in the figure represent the corresponding helicities.}
\end{figure}

However, during the past decade the handbag mechanism has become the leading theoretical approach for extracting nucleon quark and gluon  structure from exclusive reactions  such as deeply virtual Compton scattering (DVCS) and deeply virtual meson electroproduction (DVMP). In this approach the quark distributions are parameterized in terms of  generalized parton distributions (GPDs).  The GPDs contain information about the distributions of both the longitudinal momentum and the transverse position of partons in the nucleon. In the handbag mechanism  the reaction amplitude factorizes into two parts. One part describes the basic hard electroproduction process with a parton within the nucleon,  and the other - the GPD- contains the distribution of partons within the nucleon which are the result of  soft processes. While the former is reaction dependent, the latter is a universal property of nucleon structure common to the various exclusive reactions. This is schematically illustrated in Fig.~\ref{fig:handbag}. While the handbag mechanism should be most applicable at asymptotically  
large photon virtuality $Q^2$,
DVCS experiments at $Q^2$ as low as 1.5~GeV$^2$  appear to be described rather well at leading twist by the handbag mechanism, while the range of validity of leading order applicability of DVMP is not as clearly determined.
  
There are eight GPDs. Four correspond to  parton helicity conserving (chiral-even) processes,  denoted 
by $H^q$,  $\tilde H^q$,  $E^q$ and  $\tilde E^q$. 
Four correspond to parton helicity-flip (chiral-odd) processes  \cite{diehl,ji},  $H^q_T$,  $\tilde H^q_T$,  $E^q_T$ and  $\tilde E^q_T$. 
The GPDs depend on three kinematic variables: $x$, $\xi$ and $t$, where $x$ is the average parton longitudinal momentum fraction and $\xi$ (skewness) is half of the longitudinal momentum fraction transferred to  the struck parton. The skewness can be expressed in terms of the   Bjorken variable $x_B$  as
$\xi\simeq x_B/(2-x_B)$, in which $x_B=Q^2/(2p\cdot q)$,  $q$ is  the four-momentum of the virtual photon and $Q^2=-q^2$. The momentum transfer to the nucleon is $t=(p-p^\prime)^2$, where $p$ and $p^\prime$ are the initial and final four momenta of the nucleon.

 In the forward limit where $t\to 0$,  $H^q$ and  $\tilde H^q$   reduce to the parton density distributions $q(x)$ and parton helicity distributions $\Delta q(x)$ respectively. The first moments in  $x$ of the chiral-even GPDs are related to the elastic form factors of the nucleon:  the Dirac form factor   $F_1^q(t)$,  the Pauli form factor  $F_2^q(t)$,  the axial-vector form factor  $g_A^q(t)$ and the  pseudoscalar form factor $h_A^q(t)$ \cite{polyakov}.  
  
Most of the reactions studied, such as DVCS or vector meson production, are at leading order primarily  sensitive to the  chiral-even GPDs.  
Very little is known about the chiral-odd  GPDs.  $H_T^q$ becomes the transversity function $h_1^q(x)$ in the forward limit.  The chiral-odd GPDs are difficult to access since  subprocesses with a quark helicity-flip are suppressed. However, a  complete description of nucleon structure requires the knowledge of the transversity GPDs as well as chiral even GPDs.

Pseudoscalar meson electroproduction, and in particular $\pi^0$ production in the
reaction $ep\to e^\prime p^\prime \pi^0$, was identified~\cite{Ahmad:2008hp,G-K-09} as especially sensitive to the helicity-flip subprocesses. Evidence of their possible contribution to $\pi^+$ electroproduction in target spin asymmetry data  \cite{hermes-transverse} was noted in Ref.~\cite{G-K-09}. A disadvantage of   
$\pi^+$ production is that the interpretation is complicated by the dominance  of the longitudinal $\pi^+$-pole term, which is absent in $\pi^0$ production. 
 In addition, for $\pi^0$ production the structure of the amplitudes further suppresses the quark helicity conserving amplitudes relative to the  helicity-flip  
 amplitudes~\cite{G-K-09}.  On the other hand, 
$\pi^0$ cross sections  over a large kinematic range are much more difficult to obtain than for  $\pi^+$ for two reasons: First, the cross sections are much smaller than for $\pi^+$, and second, the clean detection of $\pi^0$s requires the measurement of their two decay photons.

This letter presents the results of a measurement of  $\pi^0$ electroproduction cross sections. The primary focus here is in its interpretation within the framework of the handbag model and on its sensitivity, within this framework, of accessing the quark helicity flip  GPDs.

The  handbag mechanism is schematically  illustrated in  Fig.~\ref{fig:handbag}. The reaction can be written as a linear sum of amplitudes, each of which factorizes into two processes. In the framework of Ref.~\cite{ji}:

1. A  process  in which the incident virtual photon of helicity $\mu=0,\pm 1$ interacts with a single quark within the nucleon having a momentum fraction $x+\xi /2$ and helicity $\lambda=\pm 1/2$,  to produce a meson with helicity $\mu^\prime = 0$ and a returning quark with momentum fraction $x-\xi /2$ and helicity $\lambda^\prime = \pm 1/2$,  which is absorbed to form the final nucleon. In the present study for transversely polarized photons $\lambda^\prime =-\lambda,\ \mu= \pm 1$ and $\nu^\prime=\pm \nu$.

2. Process 1 is convoluted with  a
GPD, which  encodes the  distribution of quark and gluon longitudinal momentum fractions and transverse spatial distributions  within the nucleon. 

The primary contributing GPDs in meson production for transverse photons are  $H_T$, which characterizes the quark distributions involved in nucleon helicity-flip, and   $\bar E_T (= 2\widetilde H_T + E_T) $ which characterizes the quark distributions  involved in  nucleon  non-helicity-flip processes~\cite{diehl2},\cite{Goekeler}. 
This GPD describes the density  of transversely  polarized quarks in an unpolarized nucleon~\cite{diehl2},\cite{Goekeler}. 


The relative contributions of the nucleon helicity-flip
and nucleon helicity non-flip  processes  determine the $t$ dependence of the differential cross sections.

Exclusive $\pi^0$  electroproduction was measured at Jefferson Lab with the  CLAS  large acceptance spectrometer~\cite{clas-detector} .   Cross sections were extracted over a wide range in 
$Q^2$, $t$ , $x_B$ and  $\phi_\pi$ (the azimuthal angle of the pion production plane relative to the electron scattering plane.)
The  incident electron beam energy was 5.75 GeV. The target was  liquid hydrogen of length 2.5 cm. 
The integrated luminosity  was 20 fb$^{-1}$. 
The CLAS detector consists of six identical sectors
within a toroidal magnetic field. Each sector is equipped with three layers of drift chambers to determine the trajectory of charged particles, a gas Cherenkov counter  for  electron identification, a scintillation hodoscope for time-of-flight  measurement,  and an electromagnetic calorimeter (EC) for electron identification and photon detection for angles greater than 21$^\circ $. A forward angle calorimeter  was added to the standard CLAS configuration downstream of the target for the detection of pion decay photons  in the forward direction (4.5$^\circ$ to 15$^\circ$).  A  superconducting solenoid  around  the target  was used to trap Moller electrons along  the beam axis, while permitting detection of   photons starting at  4.5$^\circ$, protons  in the range 21$^\circ$  to 60$^\circ$, and electrons from 21$^\circ$  to 45$^\circ$. 
All four final-state particles of the reaction $ep\to e^\prime p^\prime \pi^0$, $\pi^0\to \gamma\gamma$ were detected. 


The kinematic requirements for the accepted data were: $Q^2 \ge 1$ GeV$^2$,  center-of-mass energy  $W \ge 2$ GeV, and scattered electron energy $E^\prime \ge 0.8$ GeV. The corresponding range  of $x_B$ was from 0.1  to 0.58. The electrons were identified by requiring both a Cherenkov  signal and an appropriate  energy deposition in the EC calorimeter.  Protons were identified by TOF measurement. 
Geometric  cuts were applied to include only   regions of the detector with well understood acceptance and efficiency,
as well as electron and proton target vertex position cuts, to ensure well-identified events.

The photons from $\pi^0\to \gamma\gamma$ decays were detected in the electromagnetic calorimeters.
Once all final particles were identified,  the exclusive reaction $ep\to e^\prime p^\prime \pi^0$ was selected as follows:
The angle between the direction of the reconstructed  $\pi^0$s   and the missing   momentum for
$ep\to e^\prime p^\prime X$ 
had  to be less than $\ 2 ^\circ$. $3\sigma$ cuts were made on  the missing mass 
$M_X^2(ep\to e^\prime p^\prime X) = m^2_{\pi^0}$,  the missing mass 
$M_X(ep\to e^\prime \gamma\gamma X)=  M_{p}$, the missing energy $E_X(ep\to e^\prime p^\prime \pi^0)= 0$, and the invariant mass $M_{\gamma\gamma} = m_{\pi^0}$.
The background under the $\pi^0$ invariant mass peak, typically 3 to 5\%, was  subtracted using the data in the sidebands.

Corrections for the inefficiencies in track reconstruction and detector inefficiencies were applied.
The acceptance was calculated using the standard {\small GEANT3}-based CLAS Monte-Carlo simulation software.  The Monte-Carlo generator  for  exclusive $\pi^0$ electroproduction was parameterized to be consistent with  the data. 
The ratio of the number of reconstructed Monte-Carlo events to the data events was typically a factor of about 12.
Thus the statistical error introduced by the acceptance calculation was much smaller than for the  data.

The data were binned in $Q^2,x_B,t$ and $\phi_\pi$, and differential cross sections 
$d^4\sigma/dQ^2dx_Bdtd\phi_\pi$
were obtained for more than 1800 bins.

Radiative corrections  were calculated using the software package {\small EXCLURAD}~\cite{exclurad}, which had been previously developed and used for analyzing earlier  CLAS $\pi^0$ experiments. 
Radiative corrections depend on $Q^2, t, x_B$ and $\phi_\pi$. 
They vary from  5  to 10\%, depending on the kinematics.

An overall normalization factor of 1.12  was obtained from  comparing  elastic cross sections requiring $e$-$p$ coincidence, with published data. A systematic uncertainty  of $\pm 6$\% was applied to the resulting cross sections due to this correction.

Other systematic uncertainty studies included the electron, proton and photon particle identification, the variation of the
cuts on missing masses  $M_X(ep\to e^\prime \gamma\gamma X)$ and $M_X(ep\to e^\prime p^\prime X)$, missing energy, fiducial volumes,
invariant mass $M_{\gamma\gamma}$ and   radiative corrections.
The overall systematic uncertainties were  estimated at about 10\%. 

\begin{figure*}[!ht]
\begin{center}
\hspace{-0.3 in}
\includegraphics[scale=0.9]{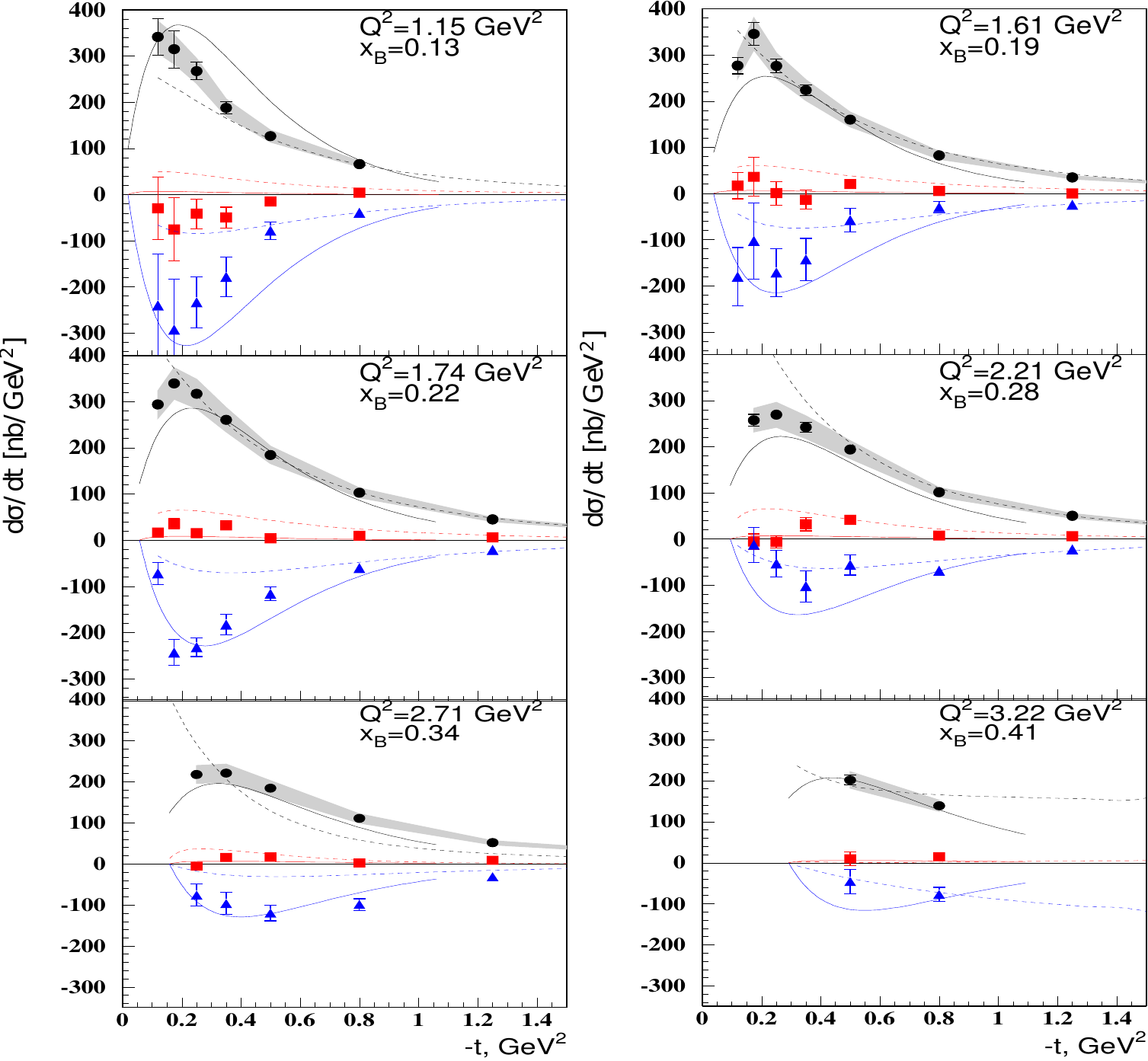}
\end{center}
\vspace{-0.35in}
\caption{ \label{fig:GK-GL} The extracted structure functions vs. $t$ for the  bins  with the best kinematic coverage and  for which there are theoretical calculations. The data and curves are as follows:  black-$\sigma_U (=\sigma_T +\epsilon \sigma_L)$,\  blue-$\sigma_{TT}$ ,  and  red-$\sigma_{LT}$. 
The shaded bands reflect the experimental systematic uncertainties.The curves are theoretical predictions produced  with the models of Refs.~\cite{GK-ps-11} (solid) and~\cite{GGL_odd_2011} (dashed).}
\end{figure*} 

The structure functions are related to the differential cross sections by~\cite{G-K-09}
\begin{eqnarray}\label{sigmatot}
\nonumber \frac{d^4\sigma}{dQ^2 dx_B dt d\phi_\pi}=\Gamma(Q^2,x_B,E)\frac{1}{2\pi} {( \sigma_{T}+\epsilon\sigma_{L}}
\\
+ \epsilon \cos 2 \phi_\pi  \sigma_{TT} + \sqrt{2\epsilon(1+\epsilon)} \cos \phi_\pi  \sigma_{LT} ).
\end{eqnarray}
\noindent The Hand convention~\cite{Hand} was adopted for the definition of the virtual photon flux factor $\Gamma$.
%
%
%
%
%
The unseparated cross section
$\sigma_U=\sigma_T+\epsilon \sigma_L$, and the interference terms 
$\sigma_{LT}$ and 
$\sigma_{TT}$  were extracted from  the  $\cos \phi_\pi$ and $\cos 2\phi_\pi$ dependences of the cross sections. 
The extracted  structure functions as functions of $-t$  are presented in Fig.~\ref{fig:GK-GL} for 6  of the  17 bins in $Q^2$ and $x_B$  bins, for which have the largest  kinematic coverage and for which there are theoretical calculations.  A recent experiment, Ref.~\cite{Hall-A-pi0}, measured $\pi^0$ cross sections in a limited kinematic range. When their results are projected to the present $Q^2$ the unseparated cross sections agree within a few percent.

The results of two GPD-based models \cite{GK-ps-11,GGL_odd_2011}  are  superimposed in Fig.~\ref{fig:GK-GL}.
The contributions from transversely polarized photons are primarily from  $H_T$  and   $\bar E_T$.    Reference~\cite{GK-ps-11} obtains the following relations:
\noindent 
\begin{eqnarray}\label{sigmat}
\sigma_{T}   =                  
\frac{4\pi \alpha_e}{2\kappa} 
\frac{\mu_\pi^2}{Q^4}
[(1-\xi^2)|\langle  H_T \rangle|^2  -  
\frac{t^\prime}{8m^2}|\langle \bar E_T \rangle |^ 2] 
\end{eqnarray}
\noindent and 
\noindent 
\begin{eqnarray}\label{sigmatt}
\sigma_{TT}= \frac{4\pi \alpha_e}{2\kappa}\frac{\mu_\pi^2}{Q^4} \frac{t^\prime}{8m^2}|\langle \bar E_T \rangle |^2.
\end{eqnarray}
\noindent Here $\kappa(Q^2,x_B)$ is a phase space factor, $t^\prime =t-t_{min}$, and the brackets $\langle  H_T \rangle$ and $\langle \bar E_T \rangle$ denote
the convolution of the elementary process with the GPDs $H_T$ and $\bar E_T$.
%



The contribution $\sigma_L$  accounts for only a small fraction in both  calculations
(typically less than  a few  percent) of the unseparated  $\sigma_T+ \epsilon \sigma_L$ in the kinematic regime under investigation. This is because  $\tilde H$  and $\tilde E$,  the GPDs which are responsible for the leading-twist structure function $\sigma_L$,  are very small. This is not the case for $\bar E_T$  and $H_T$ which contribute to $\sigma_T$ and $\sigma_{TT}$.
In addition,  the transverse cross sections are strongly enhanced by the chiral condensate through
the parameter 
$\mu_\pi=m^2_\pi/(m_u+m_d)$, where $m_u$ and $m_d$ are current quark masses \cite{G-K-09}.

 With the inclusion of the quark helicity non-conserving  chiral-odd GPDs, which  contribute primarily to $\sigma_T$ and  $\sigma_{TT}$  and, to a lesser extent $\sigma_{LT}$, the model agrees moderately well with the data. Deviations in shape become greater at smaller $t^\prime$ for the unseparated cross section $\sigma_U$. The behavior of the cross section near the threshold 
$t^\prime$ is determined by the interplay between $H_T$ and $\bar E_T$. If $\bar E_T$ dominates, the cross section becomes small as  $t^\prime \to 0$. 
For the GPDs of Ref.~\cite{GK-ps-11}  the parameterization was guided by the lattice calculation results of Ref.~\cite{Goekeler},  while Ref.~\cite{GGL_odd_2011} used a GPD Reggeized diquark-quark model to obtain the GPDs.
The results  in Fig.~\ref{fig:GK-GL} 
for  the model of Ref.~\cite{GK-ps-11} (solid curves), in which  $\bar E_T$  is dominant,  agree rather well with the data. In particular, the  structure function $\sigma_U$ begins to decrease as $-t$ becomes small, showing the effect of $\bar E_T$.    In the model of Ref.~\cite{GGL_odd_2011}(dashed curves) $H_T$ is dominant, which leads to a large rise in cross section as $-t^\prime$ becomes small. Thus, in their parameterization, the relative contribution of $\bar E_T$ to $H_T$ appears to be underestimated.
One can make a similar  conclusion from  the comparison between data and model predictions for  
$\sigma_{TT}$. This  shows the sensitivity of the measured  $\pi^0$ structure functions  for constraining  the transversity GPDs.

From  Eq.~(\ref{sigmat})  for $\sigma_T$  and Eq.~(\ref{sigmatt})  for $\sigma_{TT}$ one can conclude
that $|\sigma_{TT}|<\sigma_T<\sigma_U$.
One sees from Fig.~\ref{fig:GK-GL} that $-\sigma_{TT}$ is a sizable fraction of the unseparated cross section while $\sigma_{LT}$ is very small, which implies that contributions from transversely polarized photons play a dominant  role in the $\pi^0$ electroproduction process.

In conclusion, differential cross sections of exclusive pion electroproduction have been obtained in the few GeV region over a wide range of $Q^2, x_b$ and $t$.
While the general features of $\pi^0$ electroproduction have been described by recent Regge models~\cite{laget,kaskulov}, the focus of this letter is on the handbag mechanism in terms of quark and gluon degrees of freedom. Within the handbag interpretation, the data appear to confirm the expectation that  pseudoscalar, and in particular $\pi^0$,  electroproduction is a uniquely sensitive process to access the transversity GPDs $\bar E_T$ and $H_T$.  The measured unseparated cross section is much larger than expected from leading-twist handbag calculations.  This means that the contribution of the longitudinal cross section $\sigma_L$ is small in comparison with $\sigma_T$. The same conclusion can be made in an almost model independent way from comparison of  the  cross section $\sigma_U$, 
$\sigma_{TT}$ and $\sigma_{LT}$ \cite{Kroll_private}. 

Detailed interpretations are model dependent and quite dynamic in that they are strongly influenced by new data as they become available. In particular, calculations are in progress to compare the theoretical models  with the 
single beam spin asymmetries obtained earlier with CLAS~\cite{demasi} and longitudinal target spin asymmetries which are currently under analysis. 

In the near future new data on  $\eta$ production and ratios of $\eta$ to $\pi^0$ cross sections are expected to further constrain GPD models.
Extracting $\sigma_L$ and  $\sigma_T$ with improved statistical accuracy and performing new measurements with transversely  and longitudinally polarized targets would also be very useful.

We thank  the staff of the Accelerator and Physics Divisions at Jefferson Lab for making the experiment possible. We also thank G. Goldstein, S. Goloskokov, P. Kroll, J. M. Laget and S. Liuti  for many informative discussions and clarifications of their work, and making available the results of their calculations. 
This work was supported in part by 
the U.S. Department of Energy and National Science Foundation, 
the French Centre National de la Recherche Scientifique and Commissariat  \`a l'Energie Atomique, the French-American Cultural Exchange (FACE),
the Italian Istituto Nazionale di Fisica Nucleare, 
the Chilean Comisi\'on Nacional de Investigaci\'on Cient\'ifica y Tecnol\'ogica (CONICYT),
the National Research Foundation of Korea, 
and the UK Science and Technology Facilities
Council (STFC).
The Jefferson Science Associates (JSA) operates the Thomas Jefferson National Accelerator Facility for 
the United States Department of Energy under contract DE-AC05-06OR23177.

\end{document}